\documentstyle[preprint,aps]{revtex}
\draft
\preprint{ \begin{tabular}{l}
\hbox to\hsize{November, 1998 \hfill SNUTP 98-130,\ KIAS-P98041}\\
\hbox to\hsize{\hfill hep-ph/9811510}\\
\end{tabular} }
\tightenlines
\begin{document}

\def\beqar{\begin{eqnarray}}
\def\eeqar{\end{eqnarray}}
\def\beq{\begin{equation}}
\def\eeq{\end{equation}}
\def\tnu{\tilde{\nu}}
\def\tS{\tilde{S}}
\def\abs#1{\left|#1\right|}

\title{\Large\bf $R_F$ Parity and Almost Massless Up Quark}
\author{Jihn E. Kim$^{(a,b)}$, Bumseok Kyae$^{(a)}$, and
Jae Sik Lee$^{(b)}$ \\} 
\address{$^{(a)}$Department of Physics and Center for Theoretical
Physics, Seoul National University,\\
Seoul 151-742, Korea, and\\
$^{(b)}$School of Physics, Korea Institute for Advanced Study, 
Cheongryangri-dong, Dongdaemun-ku, Seoul 130-012, 
Korea}
\maketitle

\begin{abstract} 
We introduce a parity $R_F$ to introduce naturally small
masses for the first family members and in particular
almost massless $u$ quark toward the strong CP solution. 
We also discuss the phenomenological implications of this model 
on the proton decay and the neutrino mass.
Furthermore, it is possible to embed this $R_F$ parity to local 
$U(1)_R$ gauge symmetry. 
\end{abstract}

\newpage

The massless u-quark scenario is one of the attractive solutions
of the strong CP problem \cite{rev}, which seems to be still
alive \cite{uquark},\footnote{Note, however, that there is another
school favoring $m_u\simeq 5$~MeV \cite{leut}.} but there does not exists
a compelling model for this.\footnote{H. Georgi introduced a 
$U(1)$ gauge symmetry for this purpose \cite{georgi}.} Toward a
solution of this kind, one must introduce a symmetry which 
distinguishes the family number and is probably broken spontaneously. 
If this symmetry guarantees the masslessness of the u-quark only,
then the symmetry need not be broken. But, if the
symmetry renders other particles such as electron and d quark
massless, then it must be spontaneously broken.

The most abnormal masses in the standard model is the masses of
the first family which are $10^{-5}$ times smaller than the
electroweak symmetry breaking scale. Because of the smallness of these
masses, the radiative generation of the first family masses
were considered before, but has not led to massless u-quark \cite{barr}.

Toward the solution of the gauge hierarchy problem, supersymmetry
seems to be needed \cite{review}. In this supersymmetric scenario,
we encounter the unwanted R-parity violating terms in general.
To forbid these, the R-parity defined as $R=(-1)^{3B+L+2S}$ is assumed
to be conserved \cite{weinberg}.
 
In this paper, we formulate a theory
with naturally small electron and d-quark masses and
almost vanishing u quark mass.  Toward this purpose,
let us generalize the R-parity to $R_F$ 
so that the family information is encoded,
\begin{equation}
R_F=(-1)^{3B+L+2S}(-1)^{2IF},
\end{equation}
where $B, L, S, I$ and $F$ are the baryon number, lepton number, spin, 
weak isospin, and the first family number, respectively. Namely,
\begin{equation}
F=\delta_{f1},
\end{equation} 
where $f=1,2,3$. Then, the $R_F$ quantum numbers of chiral superfields
are
\begin{eqnarray}
&\ \ L_1\ \ \ \ L_2\ \ \ \ L_3\ \ \ \ E_1^c\ \ \ \ E_2^c\ \ \ \ E_3^c
\nonumber\\
& +1\ \ -1\ \ -1\ \ \ -1\ \ -1\ \ -1\nonumber\\
&\ \ \nonumber\\
&\ \ Q_1\ \ \ \ Q_2\ \ \ \ Q_3\ \ \ \ U^c_1\ \ \ \ U_2^c\ \ \ \ U_3^c
\ \ \ \ D_1^c\ \ \ \ D_2^c\ \ \ \ D_3^c\nonumber\\
&+1\ \ -1\ \ -1\ \ -1\ \ -1\ \ -1\ \ \ \ -1\ \ -1\ \ -1\\
&\ \ \nonumber\\
&\ \ H_1\ \ \ \ H_2\nonumber\\
& +1\ +1\nonumber
\end{eqnarray} 
where $H_1,H_2$ are the Higgs superfields with $Y=-1/2,+1/2$,
respectively. All quark singlet superfields are given $R_F=-1$
so that there is no $\lambda^{\prime\prime}$ coupling.  
The lightest $R_F=-1$ particle is a neutrino or u-quark.

The most general $d=3$ superpotential
consistent with the $R_F$ parity is
\begin{eqnarray}
W_0=&f^{(l)}_{ij}L_iE_j^cH_1(i\ne 1)+f^{(u)}_{ij}Q_iU_j^cH_2(i\ne 1)
+f^{(d)}_{ij}Q_iD^c_jH_1(i\ne 1)\nonumber\\
&+\lambda_{1jk}L_1L_jE^c_k(j\ne 1)+\lambda^\prime_{i1k}L_iQ_1D^c_k(i\ne 1)
+\lambda^\prime_{1jk}L_1Q_jD^c_k(j\ne 1),
\end{eqnarray}
which gives the following $Q_{em}=2/3$ and $Q_{em}=-1/3$ quark mass
matrices,
\begin{equation}
M^{(2/3)}=\left(\matrix{0,\ H_2^0,\ H_2^0\cr 
                        0,\ H_2^0,\ H_2^0\cr
                        0,\ H_2^0,\ H_2^0 }\right),\ \ 
M^{(-1/3)}=\left(\matrix{\tilde l_{2,3}^0,\ H_1^0\ {\rm and}\ \tilde l_1^0,
                           \ H_1^0\ {\rm and}\ \tilde l_1^0\cr
                         \tilde l_{2,3}^0,\ H_1^0\ {\rm and}\ \tilde l_1^0,
                           \ H_1^0\ {\rm and}\ \tilde l_1^0\cr
                         \tilde l_{2,3}^0,\ H_1^0\ {\rm and}\ \tilde l_1^0,
                           \ H_1^0\ {\rm and}\ \tilde l_1^0 }\right),
\end{equation}
where we have suppressed the Yukawa couplings. The rows count the
singlet anti-quarks, and the columns count the doublet quarks.
The charged lepton matrices have the same form as the $Q_{em}=-1/3$
quark mass matrices.  

It is obvious that Det$M^{(2/3)}=0$, implying the massless u-quark.
The special feature in supersymmetry is that the Higgs fields
giving masses to u- and d-quarks are different. Only, $H_2$ can
give a mass to u-quark. But d quark can obtain mass from $H_1$
and {\it also from sneutrinos.} This makes the difference between u- and
d-quarks. In the limit of vanishing sneutrino VEV's, the first
family masses are all zero, rendering a partial hierarchy of masses
among families. 

However, if this vanishing sneutrino VEV's
of the second and third family persists, the model is futile
due to the experimental facts on $m_e\ne 0$ and $m_d\ne 0$. Thus,
we require that sneutrinos of the second and/or the
third generation should develop a small vacuum expectation 
value so that d-quark and electron obtain masses. 

The $d=2$ superpotential consistent with the $R_F$ parity is
\cite{mui}
\begin{equation}
\mu H_1H_2+\mu_1L_1H_2.
\end{equation}
Thus $\tilde \nu_1=\tilde l_1^0$ develops a VEV
since the scalar potential with soft terms contains
$$
m_0^2|\tilde l_1^0|^2 + (A\mu_1\frac{v_2}{\sqrt{2}}~\tilde l_1^0+{\rm h.c.}),
$$
where $m_0$ and $A$ are of order supersymmetry breaking
scale $M_{\rm SUSY}$. However, with the fields
given in Eq.~(2), $\tilde \nu_{2,3}$ can never obtain VEV's,
and hence electron and d-quark remain massless. We must
include other fields to give VEV's to $\tilde\nu_{2,3}$.
As a minimal example, let us introduce a singlet superfield $S$ with
$Y=0$ and $R_F=-1$ and introduce a small explicit $R_F$ violating
$\epsilon^2$ term in the singlet sector superpotential
\begin{equation}
W_1=M_S S^2+\epsilon^2 S + f_S^i SL_iH_2(i\neq1) +
{\tilde \lambda_{Sijk}^{\prime\prime}\over M_P}SD_i^cD_j^cU_k^c,
\end{equation}
where $M_P$ is the Planck mass. To see the physical effects of breaking 
the $R_F$ parity, we introduced
the softly breaking $\epsilon^2$ term, which is hoped to mimic a
general feature in other $R_F$ breaking models.  
We require $|\epsilon|\ll M_S$ so
that $R_F$ breaking is soft and weak,
and $M_S$ will be constrained later. 
For simplicity, let us consider the case $f^3_S=0$ and $f^2_S
\equiv f_S$. Then, the scalar potential is described by
\beq
V=V_F+V_D+V_{\rm soft}.
\eeq
Near $(S,\tnu_2)=(0,0)$, the relevant terms are
\beqar
V_F&\simeq&\abs{M_S}^2\abs{S}^2+\frac{\abs{f_S}^2v_2^2}{2}\abs{\tnu_2}^2
+\frac{\abs{f_S}^2v_2^2}{2}\abs{S}^2
+\left[M_S\left(\epsilon^2+f_S\frac{v_2}{\sqrt{2}}~\tnu_2\right)^* S+{\rm
h.c.}\right], \\
V_D&\simeq&\frac{M_Z^2\cos2\beta}{2}\abs{\tnu_2}^2, \\
V_{\rm soft}&\simeq&m_{S}^2\abs{S}^2+m_{\tnu_2}^2\abs{\tnu_2}^2+
\left[B_S M_S S^2+B_{\epsilon}\epsilon^2 S+A_Sf_S
\frac{v_2}{\sqrt{2}}~\tnu_2~S
+{\rm h.c.}\right],
\eeqar
where $m_S$, $m_{\tnu_2}$, $B_S$, $B_{\epsilon}$, and $A_S$ are of order
supersymmetry breaking scale $M_{\rm SUSY}$.
There appear linear terms ($\epsilon^2$ terms) for $S$ in Eq.~(9)
and Eq.~(11), and hence $S$ develops a VEV. At this $\langle S\rangle$ vacuum,
$\tilde\nu_2$ contains linear terms also. Therefore,
$S$ and $\tnu_2$ fields develop VEV's in general. 

Now let us try to estimate VEV's for $S$ and $\tilde \nu_i$. For this
purpose, we impose the following four conditions to be satisfied:\\
\indent (i) Electron and d-quark obtain O(1) MeV masses,\\
\indent (ii) Neutrino mass is of order $5\times 10^{-2}$ eV \cite{SuperK},\\
\indent (iii) The u-quark mass is sufficiently small,
$\delta m_u<10^{-13}$~GeV \cite{kim}, and\\
\indent (iv) Proton does not decay too fast.
\vskip 0.2cm

We find that the VEV's of $S$ and $\tnu_i$ fields are typically given by
\beq
\langle S\rangle \sim \frac{\epsilon^2}{M_S},~~~~
\langle \tnu_i\rangle \sim \frac{f_S^i\,v_2\,\epsilon^2}{M^2_{\rm SUSY}}.
\eeq
In deriving the above VEV's, we assume $M_S \gg M_{\rm SUSY}$ to get the
neutrino mass consistent with the recent Super-Kamiokande data 
\cite{SuperK} as we will see.
With these VEV's, to give ${\cal O}(1)$ MeV masses 
to electron and d-quark, we obtain
\beq
\left(\left|\lambda_{i1j}\right|~{\rm or}~
\left|\lambda'_{i1j}\right| \right)
\cdot f_S^i
\cdot \epsilon^2 ~(i\neq 1) \sim {\cal O}(10)~{\rm GeV}^2,
\eeq
where we take $v_2\approx 10^2$ GeV and
$M_{\rm SUSY}\approx 1$ TeV.
Here, Eq.~(13) can be satisfied
for the largest $\lambda_{i1j}\ (i\ne 1)$ for the electron mass and
for the largest $\lambda^\prime_{i1j}$ for the d-quark mass.

Though the u-quark mass is zero at tree level, it can be generated 
radiatively when $S$ field has a vacuum expectation value.  
The one-loop $u$ quark mass is given by
\begin{equation}
\delta m_u \sim \sum_{i=2,3}\frac{f_{31}^{(u)}\lambda'_{i13}}{16\pi^2}
\frac{m_b f_S^{i} \langle S\rangle}{M_{\rm SUSY}}
\sim 3\times 10^{-4} ~\frac{f_{31}^{(u)}}{(M_S/{\rm GeV})}
~~{\rm GeV},
\end{equation}
where we take $m_b=5$ GeV and $M_{\rm SUSY}=1$ TeV.
Therefore, to solve the strong CP problem, we obtain \cite{kim}
\beq
\frac{f_{31}^{(u)}}{(M_S/{\rm GeV})}< 3 \times 10^{-10} .
\eeq

Let us now proceed to discuss the proton decay
rate and generation of neutrino mass,
resulting from the $R_F$ parity violation. 

The VEV of $S$ induces the conventional
$\lambda''$ couplings as follows
\beq
\lambda''_{ijk}\sim\frac{\tilde\lambda''_{Sijk}}{M_P}
\frac{\epsilon^2}{M_S}
\sim \tilde\lambda''_{Sijk} \left(\frac{\epsilon^2}{{\rm GeV}^2}\right)
\left(\frac{M_{\rm SUSY}}{M_S}\right)
\times 10^{-21}.
\eeq
The bounds on the product of $\lambda'\cdot\lambda''$ 
from the proton stability are \cite{product}
\beq
\lambda'_{11k}\cdot\lambda''_{11k}<10^{-24},~~~~~~
\lambda'_{\rm any}\cdot\lambda''_{\rm any}<10^{-9}.
\eeq
And there are somewhat model-dependent bounds on single $\lambda''$
\cite{single}.
Noting that there are no $\lambda'_{11k}$ couplings in our model, the induced
$\lambda''$ couplings are small enough to avoid the fast proton decay.

The mixing between $S$, neutralinos and neutrino 
($\nu_2$ in our example of $f^3_S=0$)
gives the following neutrino mass through the see-saw
mechanism
\beq
m_{\nu} \approx \frac{f_S^2\,v_2^2}{M_S}+
\frac{f_S^2\,\epsilon^4}{M_{\rm SUSY}\,M_S^2}
 - 2\frac{f_S^4\,v_2^2\,\epsilon^4}{M_{\rm SUSY}^3\,M_S^2}.
\eeq
The first term comes from the mixing between $S$ and neutrino.
The second term is the very well-known tree-level neutrino mass
from the mixing between neutralino and neutrino under the presence of 
conventional R-parity violation \cite{rmass}. We have neglected
$O(<\tilde\nu>^2/M_{\rm SUSY})$ which comes from the gaugino
intermediate tree level diagram. Since we are interested in
a generic bound on the coupling of $S$, inclusion of this
term would not change our conclusion very much. The last term of the above 
equation, coming from the overall mixing, can be neglected.
For this neutrino mass to be consistent with
the recent Super-Kamiokande data 
\cite{SuperK}, $\sqrt{\Delta m^2_{\rm atm}} \sim 5 \times 10^{-2}$ eV, 
the following relations are to be satisfied
\beqar
\frac{f_S^2}{M_S} &<&5 \times 10^{-15} 
\left(\frac{m_{\nu}}{ 5 \times 10^{-2}~{\rm eV}}\right)~~\frac{1}{\rm GeV},
~~~{\rm and}
\nonumber \\
\frac{f_S\,\epsilon^2}{M_S} &<&2 \times 10^{-4} 
\left(\frac{m_{\nu}}{ 5 \times 10^{-2}~{\rm eV}}\right)^{1/2}~~{\rm GeV}.
\eeqar

For example, let's think about the case that 
$f^{(u)}_{13}\sim3\times10^{-2}$ and
$\lambda$ or $\lambda'$ $\sim 10^{-3}$. In this case 
$f_S\,\epsilon^2 \sim 10^4$ GeV$^2$ from the Eq.~(13) and
$M_S>10^8$ GeV from the Eq.~(15). In this case the second condition of the 
Eq.~(19) is fulfilled.   
Let's take $M_S=10^8$ GeV. Then, from the first condition
of the Eq.~(19), $f_S<7\times10^{-4}$. 
Let's take $f_S=7\times10^{-4}$, then $\epsilon \sim 4\times 10^{3}$ GeV.
From all these,
$\langle S \rangle \sim 0.16$ GeV and 
$\langle \tnu \rangle \sim 1$ GeV.
Finally $\lambda''_{ijk}\sim 1.6\times 10^{-19}$.
We find that large enough $M_S$ are sufficient for the massless u-quark
scenario to be the solutions of the strong CP problem.

Finally, let us promote the $Z_2$ discrete symmetry group $R_F$ 
to a subgroup of a local $U(1)_R$ gauge group.
This kind of discrete gauge symmetry is considered to be beautiful,
because otherwise it is not guaranteed for the gravitational 
interaction to preserve the discrete symmetry \cite{dgauge}.
To consider the anomaly, 
the additive $U(1)_R$ charges of the fermionic fields are given by
\begin{eqnarray}
&\ \ L_1\ \ \ \ L_2\ \ \ \ L_3\ \ \ \ E_1^c\ \ \ \ E_2^c\ \ \ \ E_3^c
\nonumber\\
& \ \ 0\ \ \ \ \ \  1\ \ \ \ \ \  1\ \ \ \ \ \  1\ \ \ \ \ \  1\ \ \ \ \ \ 
1\nonumber\\
&\ \ \nonumber\\
&\ \ Q_1\ \ \ \ Q_2\ \ \ \ Q_3\ \ \ \ U^c_1\ \ \ \ U_2^c\ \ \ \ U_3^c
\ \ \ \ D_1^c\ \ \ \ D_2^c\ \ \ \ D_3^c\nonumber\\
& \ \ 0\ \ \  \ \ \ 1\ \ \ \ \ \ 1\ \ \ \ \ \ 1\ \ \ \ \ \ 1\ \ \ \ \ \ 
1\ \ \ \ \  \ 1\ \ \ \ \  \ 1\ \ \ \   \ 1\\
&\ \ \nonumber\\
&\ \ H_1\ \ \ \ H_2\ \ \ \ S \nonumber\\
& \ \ \ 0\ \ \ \ \ \  0\ \ \ \ \ \  1\nonumber
\end{eqnarray} 

If $U(1)_R$ with the above charges 
is broken to $Z_2$, we obtain the $R_F$ parity.
The anomaly problem of $U(1)_R$ was considered by Ibanez and Ross
\cite{anom}. Their anomaly free conditions can be summarized as
\begin{eqnarray}
\sum_i q_i^3&=&2m+ n, \nonumber \\
\sum_i q_i&=&2p+ q,\nonumber  \\
\sum_i q_i&=&2r,  \\
\sum_i q_i&=&2r',\nonumber
\end{eqnarray}
where $m,n,p,q,r$, and $r'$  are integers. The first condition
is for vanishing $U(1)_R^3$ anomaly, the second condition for
$U(1)_R$--graviton--graviton anomaly, the third condition for 
$U(1)_R$--$SU(2)^2_L$ anomaly and the last condition for
$U(1)_R$--$SU(3)^2_c$ anomaly.
The first and the second conditions are trivial in our case.  
The last two conditions are also satisfied in our case. However,
the fields given in Eq.~(20) alone do not cancel $U(1)_R$--$U(1)_Y^2$
and $U(1)_R^2$--$U(1)_Y$ anomalies. This problem can be solved
by introducing more singlets, which we will not specify.
Thus our choice of $U(1)_R$ charges makes it possible to embed our $R_F$ 
symmetry to a local gauge symmetry $U(1)_R$.\footnote{To 
make the u-quark mass zero, 
we could have interchanged the $U(1)_R$ charges
of $Q_1$ and $U_{1}^c$ fields. For this choice, however, 
anomaly free conditions for $U(1)_R-SU(2)^2_L$ and
$U(1)_R-SU(3)^2_c$ cannot be satisfied.}

In conclusion, we distinguished the families through a parity $R_F$
so that the first family members obtain naturally small masses. In
particular, the up quark mass turns out to be even smaller,
falling in the region of solving the strong CP problem. 

\acknowledgments
Two of us (JEK, BK) are supported in part by KOSEF, MOE through
BSRI 98-2468, and Korea Research Foundation.

\end{document}